# Fingerprints for anisotropic Kondo lattice behavior in the quasiparticle dynamics of the kagome metal Ni$_3$In

Dong–Hyeon Gim[1], Dirk Wulferding[1,2], Chulwan Lee[3], Hengbo Cui[1], Kiwan Nam[1], Myung Joon Han[3], and Kee Hoon Kim[1,4,*]

[1]*Department of Physics and Astronomy, Seoul National University, Seoul 08826, Korea*
[2] *Center for Correlated Electron Systems, Institute for Basic Science, Seoul 08826, Korea*
[3]*Department of Physics, Korea Advanced Institute of Science and Technology (KAIST), Daejeon 34141, Korea*
[4]*Institute of Applied Physics, Department of Physics and Astronomy, Seoul National University, Seoul 08826, Korea*

(Dated: 5 Sep 2023)

We present a temperature- and polarization-resolved phononic and electronic Raman scattering study in combination with the first-principles calculations on the kagome metal Ni$_3$In with anisotropic transport properties and non-Fermi liquid behavior. At temperatures below 50 K and down to 2 K, several Raman phonon modes, including particularly an interlayer shear mode, exhibit appreciable frequency and linewidth renormalization, reminiscent of the onset of the Kondo screening without an accompanying structural or magnetic phase transition. In addition, a low-energy electronic continuum observed in polarization perpendicular to the kagome planes reveals strong temperature dependence below 50 K, implying thermal depletion of incoherent quasiparticles, while the in-plane continuum remains invariant. These concomitant electronic and phononic Raman signatures suggest that Ni$_3$In undergoes an anisotropic electronic crossover from an incoherent to a coherent Kondo lattice regime below 50 K. We discuss the origin of the anisotropic incoherent-coherent crossover in association with the possible anisotropic Kondo hybridization involving localized Ni-3$d_{xz}$ flat-band electrons.

[*]E-mail:optopia@snu.ac.kr

## I.   INTRODUCTION

When a periodic lattice of localized electrons is embedded into a sea of itinerant charge carriers, the Kondo coupling referring to the dynamical interaction between localized and itinerant electrons can lead to a new ground state of composite quasiparticles, e.g. heavy fermions (HFs) [1]. Most renowned HF systems have been found in the rare-earth alloys, in which the localized *f*-orbital bands are strongly hybridized with surrounding *s-*, *p-,* or *d*-orbital bands. For more than half a century, such HF systems have been extensively studied to reveal rich Kondo physics such as metallicity below a coherence temperature, competition between Kondo coupling and magnetic interaction, unconventional superconductivity, quantum critical behavior, and so on [2].

On the other hand, it is relatively rare to find HF behavior in a material with a periodic array of $d$ electrons at the top-most energy levels. Although several $d$-electron systems were reported to show HF behavior [3–7], the itinerant character of $d$ electrons compared to $f$ electrons is generally detrimental to the formation of the localized band required for Kondo physics. One novel way of realizing a Kondo lattice system with $d$ electrons is to make use of the frustrated lattice geometry like the kagome crystal structure to construct a $d$-electron flat band (FB) representing localized $d$ electrons near its Fermi level $E_F$. Very recently, theoretical studies have indeed pointed out that the kagome metals with the FB at the chemical potential can develop the Kondo lattice effect analogous to the rare-earth HF compounds through the mechanism of the orbital-selective Mott transition [8,9].

On the experimental side, however, in spite of a surge of research activity on the kagome metals in recent years, it is rare to find a pristine kagome compound with the FB at the chemical potential without symmetry breaking. In many of the metallic kagome systems, the FB is far from the chemical potential and the other dispersive bands dominate low-energy physics. Moreover, numerous kagome materials have been found to undergo symmetry-breaking transitions such as a spin or a charge ordering due to presence of large exchange coupling or van Hove singularity near $E_F$, respectively [10,11]. One rare exception can be Ni$_3$In, which consists of $AB$-stacked nickel kagome planes and possesses a FB near $E_F$ arising from Ni-3$d_{xz}$ orbitals with the interlayer-bonding characteristic as predicted by the density-functional theory (DFT) calculations [12]. Angle-resolved photoemission spectroscopy (ARPES) on Ni$_3$In revealed the expected signatures of the FB, namely the featureless in-plane band dispersion at $E_F$, which is in sharp contrast to its significant dispersion along the $k_z$ direction [12].

Remarkably, several experimental signatures in Ni$_3$In are reminiscent of those observed in a HF system. For example, while Ni$_3$In exhibits non-Femi liquid (NFL) behavior, i.e., temperature ($T$)-linear in-plane resistivity below 50 K, it recovers Fermi liquid (FL) behavior with a large specific heat coefficient ($C_P/T$) of ~47 mJ/mol K$^2$ below 1 K without any sign of symmetry breaking in transport, magnetization, and specific heat data [12]. Ni$_3$In further exhibits an upturn in $C_P/T$, negative magnetoresistance scaling to magnetic field $B$ divided by temperature, and expansion of the FL region by application of $B$. All of these experimental features have been observed frequently in various HF compounds near magnetic quantum critical points, e.g. in CeCoIn$_5$ [13], YbRh$_2$Si$_{2-x}$Ge$_x$ [14], CeRhSn [15], and CePd$_{1-x}$Ni$_x$Al [16]. It has been thus pointed out that this transport and thermodynamic behavior in Ni$_3$In indicates a strong interplay between localized electrons and itinerant electrons [12].

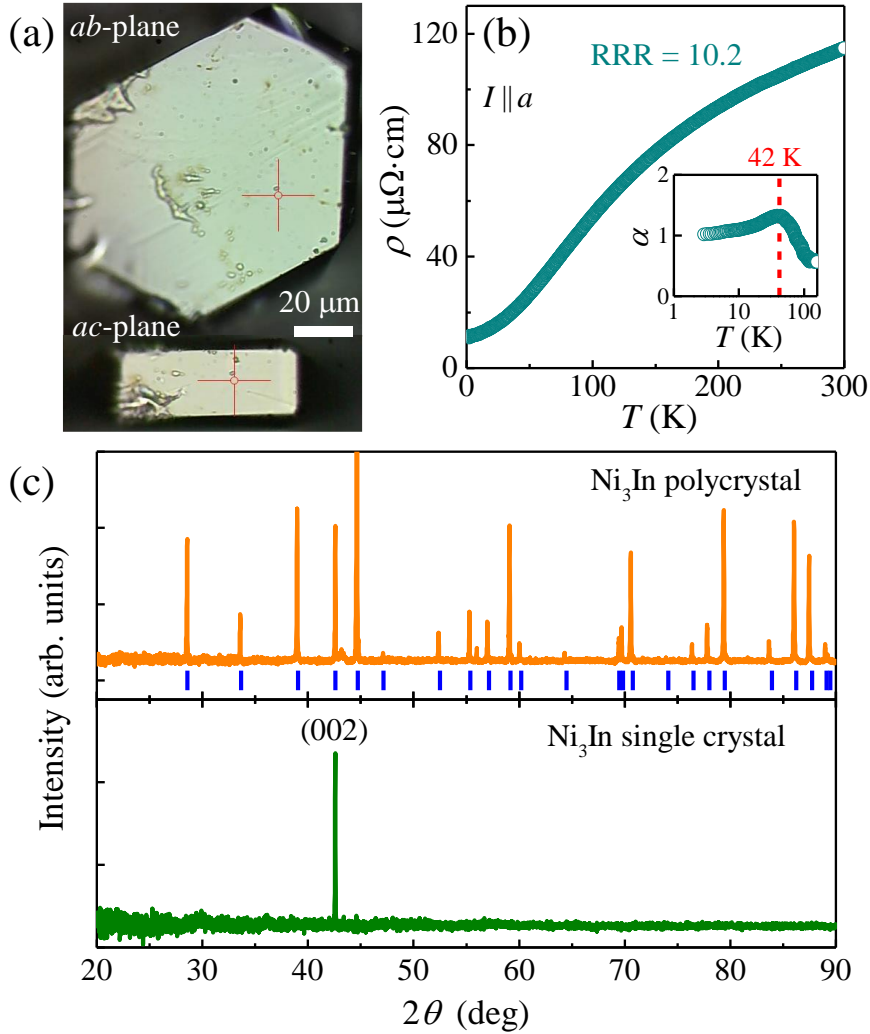

FIG. 1. (a) Optical image of *ab*- and *ac*-plane surfaces of as-grown Ni$_3$In crystals. (b) Temperature-dependent resistivity of Ni$_3$In measured in the *ab* plane, revealing a residual resistivity ratio (RRR) of 10.2. The inset shows the resistivity exponent $\alpha(T)$, with a maximum at 42 K. (c) XRD patterns of the Ni$_3$In polycrystal (top) and single crystal (bottom). The blue bars on the top panel mark the Bragg positions reported in the literature. Among the Bragg peaks, only one (002) peak is to be observed from the crystalline *ab* plane in the measured range.

Such strong interaction between localized and itinerant electrons in $Ni_3In$ may also result in distinct experimental signatures that have been often observed in archetypal Kondo lattice systems. In particular, the highly anisotropic momentum dispersion near $E_F$ as caused by the FB may facilitate anisotropic Kondo hybridization. For example, the Kadowaki-Woods ratios measured in the *ab*-plane and along the *c*-axis of $Ni_3In$ exhibit huge anisotropy with a factor larger than 300, implying an obvious distinction between the two transport channels in the FL phase [12]. It should be noted that anisotropic Kondo screening has been often observed when an *f*-electron system, being subject to the crystal field effect, exhibits highly anisotropic magnetic response or forms anisotropic electronic structure [17,18].

In order to test the possible Kondo lattice behavior in $Ni_3In$, inelastic light scattering probing both quasiparticle dynamics and phonon renormalization can be useful. In a Kondo lattice system, localized electrons hybridize with conduction electrons forming coherent quasiparticles, i.e., Kondo singlets below the coherence temperature $T^*$ [2]. Therefore, the system is expected to show strongly temperature-dependent quasiparticle excitations of incoherent quasiparticles thermally activated across the Kondo gap [19–21]; while the light scattering is reduced at low temperatures with the coherent HF state, electronic scattering increases at higher temperatures. At the same time, the Kondo hybridization is often accompanied by a phonon self-energy renormalization, resulting in frequency shifts and linewidth variations for the particular phonon modes that are receptive to the orbital overlap and related electronic hopping of the system. Therefore, the demonstration of such an incoherent-coherent crossover with temperature can provide an important clue to understand coupling between the localized and the itinerant electrons.

In this study, temperature-dependent, polarized Raman scattering has been investigated to find spectroscopic evidence for the anisotropic Kondo lattice behavior in $Ni_3In$. Quasiparticle excitations probed along both in-plane and out-plane directions and temperature-dependent phonon mode data in combination with the first principles calculations coherently show the manifestation of the incoherent-coherent crossover particularly along the out-of-plane direction below the temperature $T^* \sim 50$ K.

## II. METHODS

### A. Sample preparation

Ni$_3$In single crystals were synthesized by catalytic reactions using SnCl$_2$ as the chemical agent [12]. To produce Ni$_3$In polycrystals, vacuum-sealed nickel powder (99.9 %) and indium shots (99.99 %) were mixed in the stoichiometric ratio and reacted at 700 °C for one week in a quartz tube. The preformed polycrystals were mixed with SnCl$_2$ powder (98 %) in a mass ratio of 10:1, and then vacuum-sealed in a quartz tube. The ampoules were placed in horizontal tube furnace and heated for 4 weeks at the temperature of 560 °C. Single crystals with a typical lateral size of 150 μm (Fig. 1(a)) were found near the Ni$_3$In powder located inside the quartz tube. The composition of the as-grown crystals was confirmed using an electron probe microanalyzer. The in-plane resistivity data in Fig. 1(b) exhibit metallic behavior with a residual resistivity ratio (RRR) of 10.2, which is slightly larger than the previously reported value of ~7 [12]. Moreover, the temperature-dependent exponent $\alpha$ describing the metallic transport at low temperatures $\rho \sim T^\alpha$, as obtained by $\alpha = 1 + \frac{d}{d \ln T}\left(\ln \frac{d\rho}{dT}\right)$, becomes nearly 1 (see the inset of Fig. 1(b)). This finding clearly shows that the system exhibits the NFL behavior, being consistent with a previous result [12].

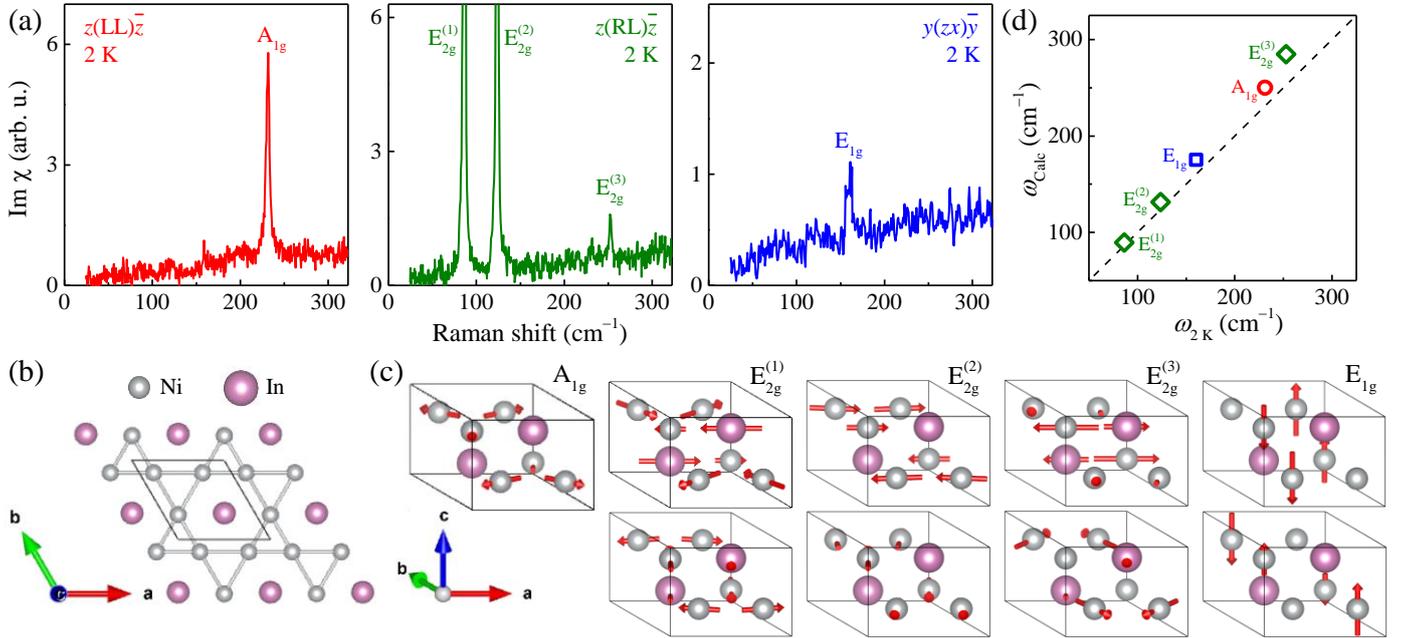

FIG. 2. (a) Raman spectra measured at 2 K in three different scattering geometries. (b) Illustrations of the kagome network of nickel atoms (gray) in each Ni$_3$In layer. (c) Atomic displacement patterns (red arrows) of the allowed Raman modes plotted in a unit cell (black lines). (d) Comparison of the calculated phonon frequencies with experimental results at 2 K. The horizontal and vertical axes represent the frequencies experimentally measured at 2 K and frequencies obtained from the DFT calculations, respectively. The dashed linear line represents an expected trace when the experimental and calculated phonon energies coincide. Red, green, and blue symbols represent the A$_{1g}$, E$_{2g}$, and E$_{1g}$ modes, respectively, being consistent with the color scheme in (a).

The X-ray diffraction (XRD) pattern of single crystals measured on the [00$l$] plane exhibits a sharp single peak, in accord with the Ni$_3$In (002) peak upon comparison with the XRD data of Ni$_3$In polycrystals (Fig. 1(c)). Five Raman peaks are observed from the grown single crystals (Fig. 2(a)), consistent with the Ni$_3$In crystal symmetry (point group $D_{6h}$) having five Raman-active modes (A$_{1g}$ + 3 × E$_{2g}$ + E$_{1g}$) with the Raman tensors;

$$A_{1g} = \begin{pmatrix} a & 0 & 0 \\ 0 & a & 0 \\ 0 & 0 & b \end{pmatrix}, \quad E_{1g} = \begin{pmatrix} 0 & 0 & 0 \\ 0 & 0 & c \\ 0 & c & 0 \end{pmatrix}, \begin{pmatrix} 0 & 0 & -c \\ 0 & 0 & 0 \\ -c & 0 & 0 \end{pmatrix}, \quad E_{2g} = \begin{pmatrix} d & 0 & 0 \\ 0 & d & 0 \\ 0 & 0 & 0 \end{pmatrix}, \begin{pmatrix} 0 & -d & 0 \\ -d & 0 & 0 \\ 0 & 0 & 0 \end{pmatrix}.$$

The atomic displacement pattern of each Raman mode is illustrated in Fig. 2(c). The three different

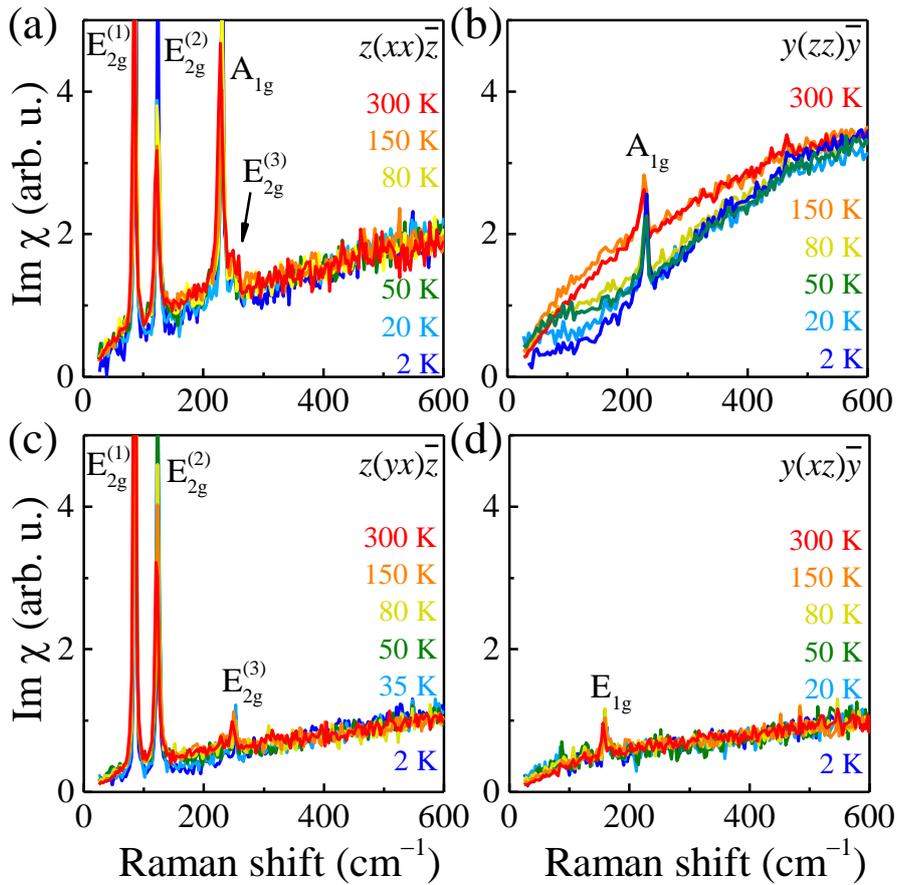

FIG. 3. Temperature-dependent Bose-corrected Raman response Im χ in the four linearly polarized scattering configurations as displayed in (a)-(d).

$E_{2g}$ modes are denoted by $E_{2g}^{(1)}$, $E_{2g}^{(2)}$, and $E_{2g}^{(3)}$ at increasing frequencies. These measurement results unambiguously characterize that the as-grown crystals comprise the *AB*-stacked $Ni_3In$ kagome nets, as drawn in Fig. 2(b).

### B. Raman scattering measurements

Raman spectra were collected in the spectral range of 25 to 600 cm$^{-1}$ using a spectrometer (TriVista™, Princeton Instruments) with a 561-nm laser and a CCD camera (PyLoN eXcellon™). The incident laser was focused on the sample surfaces with a beam spot diameter of approximately 10 μm, with a power of less than 0.3 mW to minimize the heating effect, which was estimated to be smaller than 5 K. Samples mounted on the cold-finger of a liquid He-flow cryostat were cooled down to the base temperature of 2 K and subsequently warmed up to measure the temperature-dependent spectra after waiting several hours to reach the thermal equilibrium. The Raman intensities $S(\omega,T)$ of the $Ni_3In$ crystals were Bose-corrected to obtain the Raman response $\text{Im}\,\chi(\omega,T) = S(\omega,T)/[1 + n_B(\omega,T)]$ with a Bose factor of $n_B(\omega,T) = [e^{\hbar\omega/k_BT} - 1]^{-1}$. Figure 3 summarizes the temperature-dependent Raman responses in four representative polarization and scattering configurations.

### C. First-principles calculations

First-principles calculations were carried out to predict the phonon frequencies. To calculate the electronic structure and phonon frequencies (Fig. 4), DFT calculations were performed using the

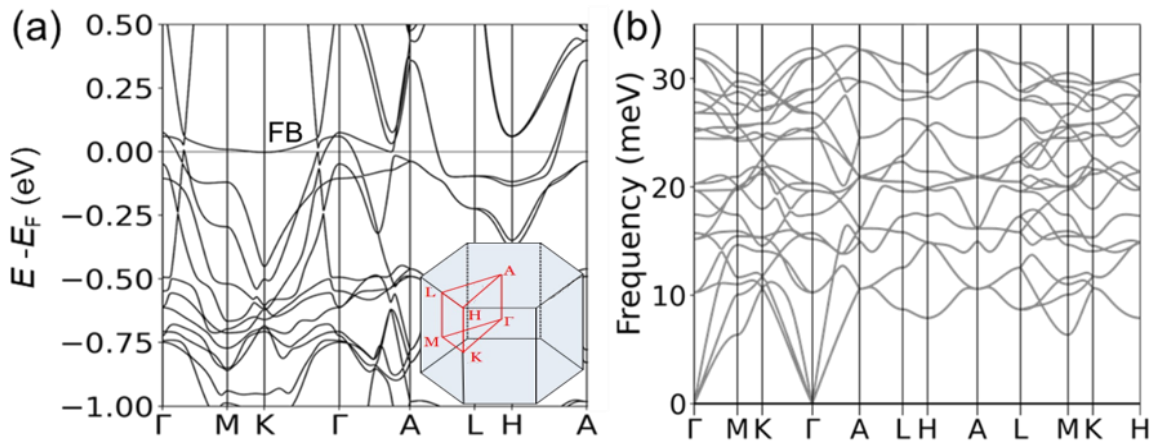

FIG. 4. (a) Electronic structure and (b) phonon energy dispersion of $Ni_3In$ calculated along the high-symmetry lines. The inset in (a) represents the Brillouin Zone of $Ni_3In$, with the high-symmetry lines and points marked in red.

Quantum Espresso software package [22]. The projector augmented wave method within fully relativistic pseudopotential scheme was adopted. Perdue-Burke-Ernzerhof parameterized exchange-correlation functional was used for electronic structure calculation. The 100 Ry of planewave energy cutoff and the 12×12×12 Monkhorst-Pack k mesh gave the well-converged results. 2×2×2 q mesh in the density functional perturbation theory was taken for obtaining the phonon dispersion in Fig. 4(b), which further shows the dynamical stability of this material. By the first-principles calculations, five Raman phonon modes are well identified at Γ point, of which frequencies are in good agreement with our Raman data (Fig. 2(d)).

## III. RESULTS

### A. Phonon anomalies

Out of those five phonon modes, the three $E_{2g}^{(1)}$, $E_{2g}^{(2)}$, and $A_{1g}$ modes, which show most conspicuous intensities, are fitted by individual Lorentzian lines to extract the temperature-dependence of phonon frequencies $\omega(T)$ and linewidths $\Gamma(T)$. Figure 5 summarizes the analysis results. Above 50 K, their temperature dependence can be largely described by an anharmonic decay model, $\omega(T) = \omega_0 - A\,[1 + 2n_B(\omega_0/2, T)] - B\,[1 + 3n_B(\omega_0/3, T) + 3n_B^2(\omega_0/3, T)]$ and $\Gamma(T) = \Gamma_0 + C\,[1 + 2n_B(\omega_0/2, T)] + D\,[1 + 3n_B(\omega_0/3, T) + 3n_B^2(\omega_0/3, T)]$, which describes the decay of an optical phonon to different lower-energy phonons in an anharmonic lattice [23,24]. Here, $\omega_0$ and $\Gamma_0$ are the bare phonon frequency and linewidth, respectively, and $A$, $B$, $C$, and $D$ are non-negative constants.

As notable in Fig. 5(a), the $E_{2g}^{(2)}$ mode deviates from the conventional hardening of optical phonon modes as predicted by the anharmonic effect and shows anomalous softening below 50 K. Namely, the maximum phonon frequency is reached around 50 K as marked by the dashed gray line in Fig. 5(a). This $E_{2g}^{(2)}$ mode corresponds to the antiphase interlayer shearing motion between the nickel kagome layers (Fig. 2(c)). The $A_{1g}$ mode representing the intralayer breathing of the trigonal nickel plaquettes also exhibits softening below ~ 50 K. In addition to those phonon softening, the linewidth of the $E_{2g}^{(2)}$ shearing mode exhibits a sharp decrease below 50 K, clearly deviating from the trajectory predicted by the anharmonic decay model (Fig. 5(c)). The linewidths of the $A_{1g}$ and $E_{2g}^{(1)}$ modes continue to decrease without saturation at low temperatures as well. The clear phonon softening and linewidth reduction constitute two pieces of evidence that nonphononic contributions to phonon self-energies are dominating below 50 K. Since the phonon energy dispersion calculated within the hexagonal phase (Fig. 4(b)) shows the dynamic stability of the crystal structure without an imaginary phonon frequency, it is unlikely that Ni$_3$In undergoes a long-range structural modulation driven e.g. by a charge density wave. We also emphasize that Ni$_3$In does not exhibit any sign of symmetry breaking around the anomaly temperature 50 K [12]. Therefore, neither a magnetic or structural long-range order nor the conventional anharmonic effects can account for the anomalous renormalization of phonon self-energies in Ni$_3$In.

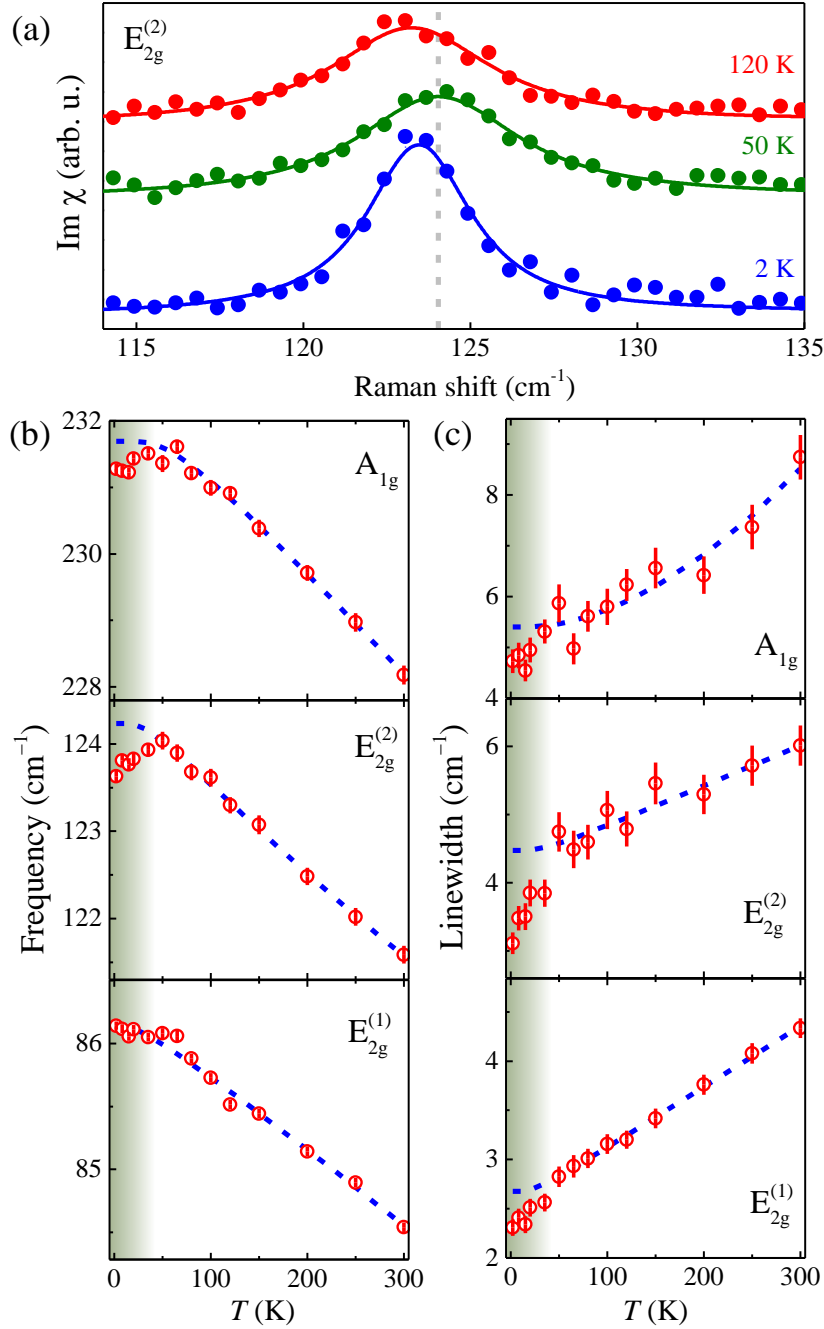

FIG. 5. (a) The $E_{2g}^{(2)}$ phonon peak at representative temperatures. Dots and solid lines are the measured data points and fitted Lorentzian profiles, respectively. The vertical gray dashed line represents the frequency of the phonon at 50 K. Temperature dependence of the center frequencies (b) and the full widths at half maxima (c), for the three Raman modes as obtained from the Lorentzian fitting. Fitted results from an anharmonic decay model are represented by the blue dashed lines, and the shaded region emphasizes a temperature window where resistivity exhibits the NFL behavior.

It should be emphasized that such a renormalization of phonon mode frequencies deviating from the anharmonic decay model has been frequently observed below $T^*$ in numerous HF compounds, e.g. CeCoIn$_5$ [19,25], Yb$_{14}$MnSb$_{11}$ [20], CeRh$_6$Ge$_4$ [21], URu$_2$Si$_2$ [26], when they enter from an incoherent high temperature regime into a coherent transport regime, featuring either FL or NFL transport behavior. These phonon frequency anomalies reported in those HF compounds are on the order or below 1 cm$^{-1}$, consistent with our observations. Moreover, as has been observed in the archetypal HF compound CeCoIn$_5$ [19,25], noticeable decrease in the phonon relaxation rates below $T^*$ indicates that concomitant with the Kondo screening, there exists an abrupt reduction in the scattering processes in the lattice. Therefore, our observation of phonon softening and linewidth reduction represents a diagnostic feature that the Kondo screening starts to occur gradually at $T \sim 50$ K, below which the coherent transport and the NFL behavior develop in Ni$_3$In.

The renowned model incorporating the coupling between itinerant charge carriers and periodic localized electrons is the Kondo lattice Hamiltonian. According to the mean-field theory of the Kondo lattice model [20], $\omega^2 - \omega_0^2$ ($\omega_0$ is a bare frequency) should be generally proportional to $\frac{d^2}{dx^2}(J_K \cdot N_K^2)$, where $J_K$ is the coupling strength between itinerant and localized electrons, $N_K$ is the density of the composite coherent quasiparticles, and $x$ is the atomic displacement of a particular phonon mode. Therefore, a particular phonon mode of the system can show a conspicuous temperature-dependent shift proportional to $N_K$, i.e., the number of the coherent quasiparticles. The observations of clear softening in both A$_{1g}$ and E$_{2g}^{(2)}$ phonon modes, and a weak variation of E$_{2g}^{(1)}$ mode frequency in Fig. 5(b) seem to be consistent with the increased $N_K$ below $T^* \sim 50$ K, i.e., the formation of coherent metallic states of the Kondo lattice system. The dramatic linewidth reduction as found in all of $E_{2g}^{(2)}$, $E_{2g}^{(1)}$, and $A_{1g}$ mode below $T^* \sim 50$ K implies that the phonon lifetime has been increased with the development of coherent metallic states, as similarly found in $f$-electron HF systems [19,25].

It should be emphasized that in Fig. 5(b) the $E_{2g}^{(2)}$ mode, representing the interlayer shearing motion, shows the most pronounced softening and sharpening below 50 K. This indicates that the product of Kondo interaction and quasiparticle density $(J_K \cdot N_K^2)$ should vary most sensitively with the atomic displacements $x$ relevant to the interlayer shearing motion, resulting in a large energy renormalization $\omega^2 - \omega_0^2 \sim \frac{d^2}{dx^2}(J_K \cdot N_K^2)$. While other Raman modes severely distort the kagome plane, the $E_{2g}^{(2)}$ mode particularly leaves each nickel trigonal plaquette almost intact but mainly alters the orbital overlap between nickel atoms located at neighboring kagome planes (Fig. 2(c)). Therefore, it is inferred that the interlayer overlap of the Ni-3$d$ orbitals is closely associated with the realization of the coherent metallic state in Ni$_3$In below $T^* \sim 50$ K.

### B. Electronic quasielastic scattering

Another salient feature that supports the Kondo lattice behavior is the increase in incoherent quasiparticles at high temperatures that are thermally excited across the potential Kondo gap [19–21]. It is well known that the low-energy continuum in the electronic Raman response can sensitively reflect the quasi-elastic scattering of incoherent charge carriers. [27,28]. Various Kondo lattice compounds such as URu$_2$Si$_2$ [29], CeCoIn$_5$ [25], and SmB$_6$ [30] have indeed exhibited such thermally enhanced low-energy electronic Raman scattering due to the thermal activation of incoherent quasiparticles. Quite similar to those representative Kondo lattice systems, the Raman response in the $y(zz)\bar{y}$ configuration exhibits conspicuous enhancement with increase of temperature (Fig. 3(b)). As the phonons are narrower than 10 cm$^{-1}$, the general spectral features of the broad-energy continuum background are clearly distinguishable from the phonon peaks. Note that the low-energy spectral weight below 100 cm$^{-1}$ exhibits a steep increase up to 50 K and nearly saturates at higher temperatures.

In sharp contrast, the Raman continua in the $z(xx)\bar{z}$, $z(yx)\bar{z}$, and $y(xz)\bar{y}$ configurations (Figs. 3(a), (c), and (d), respectively), for which the incident or scattered light polarization is within the kagome plane, do not exhibit significant differences with variation of temperature. Such distinct anisotropy in the low-energy thermal continuum appearing exclusively in the $y(zz)\bar{y}$ spectra is further verified by the parallel-polarized Raman response at 300 K in the $y(\phi\phi)\bar{y}$ configuration measured within the *ac* plane by varying the polarization angle $\phi$ with respect to the *x*-axis. In addition to the sharp $E_{2g}$ and $A_{1g}$ phonon lines located around 100 and 220 cm$^{-1}$, the contour plot of the Raman response with variation of $\phi$ clearly reveals the continuum in a broad energy range below 400 cm$^{-1}$ with the anisotropic polarization dependence (Fig. 6(a)).

To estimate the spectral weight $I(\phi)$ of the electronic Raman continuum at each $\phi$ value, the Raman response was integrated over the low-energy range below 100 cm$^{-1}$ after removing the sharp phonon lines. The polar map $I(\phi)$ (red dots in Fig. 6(b)) shows the increasing tendency towards the *z* direction ($\phi=90°$ or $270°$) characterized by a sinusoidal function $I(\phi) \sim \sin^2\phi$ (blue solid line). It is further confirmed that $I(\phi) \sim \sin^2\phi$ behavior is still valid even if the integrated energy range is expanded up to 400 cm$^{-1}$. Such strong anisotropy in the low-energy continuum as well as its steep thermal enhancement up to 50 K is hardly explainable by the effect of acoustic phonon density of states (DOS) because the acoustic phonon dispersion exhibits nearly isotropic behavior along both in-plane and out-plane directions (Fig. 3(b); see the discussions in Sec. IV).

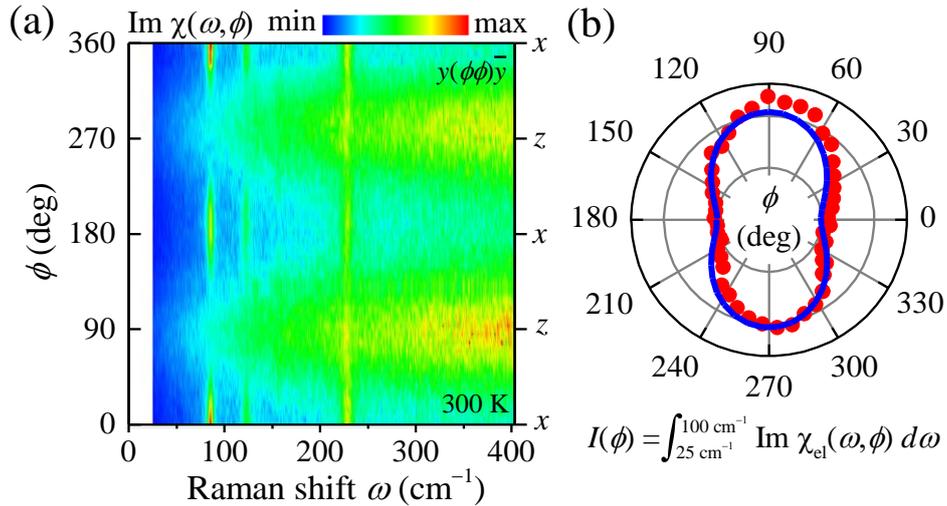

FIG. 6. (a) Polarization dependence of the $y(\phi\phi)\bar{y}$ Raman response at 300 K, in which $\phi$ denotes the angle with respect to the *x* axis. (b) The red dots indicate the angular dependence of the electronic spectral weight integrated over the low-energy range below 100 cm$^{-1}$ in (a). The blue curve represents a sinusoidal fitting function (see the text).

The steep depletion of the low-energy continuum at low temperatures below 50 K can in fact be explained by reduced thermal population of incoherent quasiparticles across a Kondo gap below 50 K. At low temperatures, the Kondo gap opening via hybridization can generate interband particle-hole excitations if the lower band is occupied. Once the lower band electrons are thermally excited across the gap at high temperatures, the thermal quasiparticles occupying the upper band should suppress the interband photoexcitation, whereas the incoherent flow of the thermally activated electrons should increase the quasielastic scattering. Therefore, the thermal activation of the incoherent quasiparticles can result in a transfer of the interband spectral weight to the low-energy continuum in the Raman response. A previous dynamical mean field theory has successfully corroborated the spectral weight transfer in the Raman response [31].

The thermal evolution of the quasielastic continuum with strong polarization dependence (Fig. 3(b)) points out that such spectral weight transfer occurs particularly along the out-of-plane direction. Namely, by the anisotropic Kondo hybridization, a coherent metallic state becomes predominant along the *c*-axis whereas the Kondo screening is mitigated in the *ab*-plane. Indeed, a previous study [12] reported more metallic transport behavior along the *c*-axis with smaller resistivity and larger RRR than in the *ab*-plane, as well as the much larger Kadowaki-Woods ratio in the *ab*-plane than that along the *c*-axis by a factor of ~ 300. Therefore, such drastic anisotropy in the transport behavior is indicative of a much higher electron-electron scattering rate in the *ab*-plane, being consistent with the anisotropic Kondo effect scenario.

It is likely that the anisotropic quasiparticle evolution originates from the anisotropic nature of the FB, which serves the role of localized electrons before forming a hybridized metallic band by the many-body Kondo effect. The anisotropic dispersion of the FB, becoming particularly flat in the $k_z = 0$ plane but dispersing significantly along the $k_z$ direction (Fig. 4(a)), arises from interlayer bonding of Ni-3$d_{xz}$ orbitals which do not overlap with intralayer neighbors but considerably with other $d_{xz}$ orbitals located in different layers [12]. Furthermore, because the (*ij*)-polarized quasielastic Raman response is weighted by an even power of an electron band curvature $\partial^2\varepsilon/\partial k_i \partial k_j$ ($i,j = x,y,z$) [27], our observations of prominent quasielastic Raman signal along the $z$-axis (Figs. 3 and 6) are consistent with the FB carrying larger out-of-plane band curvature $|\partial^2\varepsilon/\partial k_z^2|$ than the in-plane curvatures e.g. $|\partial^2\varepsilon/\partial k_x^2|$.

## IV. DISCUSSIONS

Given the absence of any sign of magnetic instability in the Ni$_3$In magnetic field-dependent transport, susceptibility, and specific heat data at 2 – 300 K [12], the characteristic low-temperature evolution of the Ni$_3$In Raman response is unlikely to be associated with a spin freezing/glass order. In addition, putative magnetic quantum fluctuations *increasing* toward zero temperature which perhaps have been observed as an upturn in *C*/*T* data [12] cannot explain the *reduction* of both low-energy Raman continuum (Fig. 3(b)) and phonon relaxation rates (Fig. 5(c)) below 50 K . To the best of our knowledge, no magnetic fluctuation mechanism without a phase transition can drive such a systematic phonon renormalization below 50 K accompanied by the reduction in fluctuations. Instead, the low-energy spectral weight is most likely incoherent quasiparticles which become coherent at 50 K concomitant with the lattice renormalization via the electronic incoherent-coherent crossover.

We also remark that the thermal evolution of the broad continuum below 50 K cannot result from anharmonic lattice effects. One of the most direct fingerprints of increased lattice anharmonicity at elevated temperatures is the occurrence of higher-order phonon scattering which can show up in Raman spectra in the form of weighted phonon DOS. However, this multi-phonon scattering cannot explain the background continuum observed from Ni$_3$In Raman spectra. As confirmed by the phonon dispersion calculation in Fig. 4(b), the acoustic phonon and optical phonon branches terminate below 100 and 280 cm$^{-1}$ respectively, whereas the observed continua in Fig. 3 extend to 600 cm$^{-1}$ well beyond the phonon frequencies. If the continuous acoustic phonon DOS dominates the Raman response, it should still produce rather characteristic features across a much narrower frequency range than the observed nearly featureless continuum. Furthermore, all of the continua display a monotonic frequency dependence with a gentle slope.

This is hard to expect from a phonon DOS spectrum, because the structured shape of DOS continua often contains multiple band components which should be much more complicated than the smooth curves observed in the Ni$_3$In Raman spectra.

Instead of higher phonon scattering processes, one might alternatively consider incoherent electrons scattering off acoustic phonons as the source of the observed electronic continuum. Such phonon scattering may be reflected in the temperature dependence of the spectral continuum below the frequency of 80 cm$^{-1}$, because energy transfer by the acoustic phonons is available up to 10 meV ≈ 80 cm$^{-1}$ (Fig. 4(b)). The low-frequency spectral weight of the continuum below 100 cm$^{-1}$ increases steeply up to 50 K and remains temperature-independent above that temperature (Fig. 3(b)). If electron scattering by acoustic phonons dominated the spectra, such quasielastic response should keep increasing for $T \geq 100$ K, since the Debye temperature $\Theta_D$ is about 116 K = 10 meV (below which the acoustic phonon branches have linear dispersions; see Fig. 4(b)). This electron-phonon scattering in an anharmonic metal therefore does not describe the observed thermal enhancement of the low-energy Raman continuum in Fig. 3(b), which in fact saturates at 50 K well below $\Theta_D > 100$ K. Therefore, we conclude that the temperature dependence, spectral features, and strong polarization anisotropy of the observed Raman continuum do not represent the increased anharmonicity in Ni$_3$In.

It is noteworthy that, even though the Kondo hybridization below $T^*$ can in principle result in a spectral weight redistribution of the electronic scattering from low to high energies, only the depletion of the low-energy spectral weight is identified in the $y(zz)\bar{y}$ spectra (Fig. 3(b)). Various Kondo insulators, e.g. SmB$_6$ [30], indeed exhibit the expected spectral weight transfer in their electronic scattering below $T^*$. On the contrary, in metallic HF compounds such as URu$_2$Si$_2$ [29], the gain in the interband spectral weight is hardly detectable from the Raman spectra, as in the current Ni$_3$In case. To understand such distinction between the insulating and metallic Kondo lattice systems, it is instructive to consider the Kondo hybridization of two simple parabolic and flat bands crossing at a finite energy $E^*$. Since the interband transitions observed in Raman experiments using laser light within the visible regime (500 – 700 nm) mainly detect direct gaps, the occupancy of the lower hybridized band determines whether the direct interband transitions can be measured after the Kondo hybridization. In Kondo insulators, because $E_F$ lies inside the Kondo gap, the fully occupied lower band can constitute an interband transition peak in a Raman spectrum with a single direct gap energy.

On the other hand, in a metallic HF case, $E_F$ falls within the bandwidth of the lower band, usually below $E^*$ so that the Fermi surface acquires heavy mass, thereby the occupied states might not have a single well-defined direct gap energy. In this case, high-energy spectral gain in the Raman response becomes faint over a much broader energy range rather than being detected as one sharp photoexcitation peak. Therefore, the details of the band occupancy and the lack of a direct gap after the hybridization might have suppressed the interband Raman scattering in Ni$_3$In below $T^*$. Moreover, in real materials, the light-matter interaction and density-density correlations become affected by screening effects and also by many-body interactions [27]. The Kondo hybridization may drastically alter the characters of the Fermi surfaces and, as demonstrated in this study, the phonons. Therefore, the dielectric property of the system is expected to change abruptly below $T^*$, and consequently, the spectral weight redistribution in Kondo lattice systems may not obey the sum rule that might hold in the non-interacting case.

We now discuss the possible origin of the anisotropic Kondo hybridization in Ni$_3$In. Previously, the anisotropic Kondo hybridization of localized and itinerant electrons has been realized in Ce-based compounds such as (Ce,La)Al$_3$ [17] and Ce(Co,Rh,Ir)In$_5$ [18], where the localized Ce orbitals subject to the crystal field effect result in the anisotropic hybridization. In the present Ni$_3$In case, the variations in the interlayer shearing mode ($E_{2g}^{(2)}$ phonon) signaling coherent quasiparticles (Fig. 5) as well as the suppression of incoherent light scattering have indeed appeared along the $c$-axis (Fig. 3(b)). This observation unequivocally implies that the coherent metallic band is well formed along the $c$-axis only. Intriguingly, as seen in the DFT calculation results of Fig. 4 (a), the FB in Ni$_3$In has flat energy dispersion particularly on the Γ-M-K-Γ plane (parallel to the $ab$ plane) in the momentum space, while it clearly disperses along the Γ-A line (perpendicular to the kagome planes). This DFT result has been confirmed by a previous ARPES measurement [12]. According to the orbital-projected DFT calculations [12], the FB comprises Ni-$3d_{xz}$ orbitals, and each Ni-$3d_{xz}$ orbital has considerable overlap with its interlayer neighbors along the *c*-axis, while the electrons become localized within the kagome plane due to reduced orbital overlap within in-plane trigonal plaquettes. Therefore, such a contrasting orbital overlap for the two crystallographic directions should be attributed to the anisotropic band dispersion of the FB in Ni$_3$In.

The anisotropic band renormalization in Ni$_3$In should arise from the anisotropic FB dispersion linked to the interlayer bonding character of Ni-$3d_{xz}$ orbitals. The critical role of the interlayer Ni-$3d_{xz}$ orbital bonding in the crossover is further evinced by the salient renormalization of the interlayer shearing mode ($E_{2g}^{(2)}$ phonon), in which the atom motions primarily alter the distance between interlayer neighbors. As a consequence, the increasing quasiparticle density below $T^*$ is expected to enhance both $d_{xz}$ orbital overlap and thus the metallic bonding along the interlayer direction, thereby affecting the lattice dynamics and the transport properties of Ni$_3$In at low temperatures. This also suggests that the nature of interlayer bonding should play a pivotal role in determining the electronic properties of other kagome metals as well.

## V. CONCLUSIONS

In summary, we have investigated the polarization- and temperature-dependent Raman response of high-quality Ni$_3$In crystals. We have discovered phonon anomalies and thermal gain of an anisotropic electronic continuum; the former implies that the population of the coherent quasiparticles below $T^* \sim 50$ K occurs by the hybridization of the localized flat band with other dispersive bands while the latter represents the thermal activation of the incoherent quasiparticles. Based on our experimental findings, we provide a plausible interpretation that Ni$_3$In can be viewed as a metallic Kondo lattice system with anisotropic Kondo coupling, which originates from the anisotropic dispersion of a 3*d*-electron flat band.

**Acknowledgements**

This work was supported by NRF of South Korea funded by Ministry of Science and ICT (2019R1A2C2090648, 2022H1D3A3A01077468) and core facility program funded by Ministry of Education (2021R1A6C101B418). D.W. was supported by Institute for Basic Science in Korea (IBS-R009-Y3). C.L. and M.J.H. were supported by NRF grant funded by Korean government (MSIT) (2021R1A2C1009303, 2018M3D1A1058754).

———————————